\documentclass[letterpaper]{article} % DO NOT CHANGE THIS
\usepackage{aaai25}  % DO NOT CHANGE THIS
\usepackage{times}  % DO NOT CHANGE THIS
\usepackage{helvet}  % DO NOT CHANGE THIS
\usepackage{courier}  % DO NOT CHANGE THIS
\usepackage[hyphens]{url}  % DO NOT CHANGE THIS
\usepackage{graphicx} % DO NOT CHANGE THIS
\usepackage{natbib}  % DO NOT CHANGE THIS AND DO NOT ADD ANY OPTIONS TO IT
\usepackage{caption} % DO NOT CHANGE THIS AND DO NOT ADD ANY OPTIONS TO IT
\usepackage{algorithm}
\usepackage{multirow}
\usepackage{amsmath}
\usepackage{booktabs}
\usepackage{xcolor}
\usepackage{graphicx}
\usepackage{amsmath}
\usepackage{amssymb}
\usepackage{booktabs}
 \usepackage{algpseudocode}
 \usepackage{algorithmicx}
\usepackage{algorithm}
\usepackage{graphicx}
\usepackage{epstopdf}
\usepackage{enumitem}
\usepackage{times}
\usepackage{helvet}
\usepackage{courier}
\usepackage{newfloat}
\usepackage{listings}
\DeclareCaptionStyle{ruled}{labelfont=normalfont,labelsep=colon,strut=off} % DO NOT CHANGE THIS
\floatstyle{ruled}
\newfloat{listing}{tb}{lst}{}
\floatname{listing}{Listing}
\title{Dynamic Dual-level Defense Routing for Continual Adversarial Training}
\author {
    Wenxuan Wang\textsuperscript{\rm 1},
    Chenglei Wang\textsuperscript{\rm 1},
    Xuelin Qian\textsuperscript{\rm 2}
}
\affiliations {
    \textsuperscript{\rm 1}School of Computer Science, Northwestern Polytechnical University,National Engineering Laboratory for Integrated Aero-Space-Ground-Ocean Big Data Application Technology\\
    \textsuperscript{\rm 2}School of Automation, Northwestern Polytechnical University,the BRain and Artiﬁcial INtelligence Lab\\
    wxwang@nwpu.edu.cn, 1474515006wcl@gmail.com, xlqian@nwpu.edu.cn
}
\usepackage{bibentry}
\begin{document}

\maketitle

\begin{abstract}
As adversarial attacks continue to evolve, defense models face the risk of recurrent vulnerabilities, underscoring the importance of continuous adversarial training (CAT). 
Existing CAT approaches typically balance decision boundaries by either data replay or optimization strategy to constrain shared model parameters.
However, due to the diverse and aggressive nature of adversarial examples, these methods suffer from catastrophic forgetting of previous defense knowledge after continual learning.
In this paper, we propose a novel framework, called Dual-level Defense Routing (DDeR), that can autonomously select appropriate routers to integrate specific defense experts, thereby adapting to evolving adversarial attacks. 
Concretely, the first-level defense routing comprises multiple defense experts and routers, with each router dynamically selecting and combining suitable experts to process attacked features. 
Routers are independently incremented as continuous adversarial training progresses, and their selections are guided by an Adversarial Sentinel Network (ASN) in the second-level defense routing.
To compensate for the inability to test due to the independence of routers, we further present a Pseudo-task Substitution Training (PST) strategy, which leverages distributional discrepancy in data to facilitate inter-router communication without storing historical data. Extensive experiments demonstrate that DDeR achieves superior continuous defense performance and classification accuracy compared to existing methods.
\end{abstract}


\section{Introduction}

Deep neural networks have achieved remarkable success in various computer vision tasks, yet their vulnerability to adversarial attacks~\cite{madry2017towards} poses a significant challenge to their reliability and security. By introducing imperceptible perturbations to inputs, attackers can manipulate or induce the model to make incorrect predictions. To mitigate this risk, adversarial training has been proposed and developed as one of the most effective defense strategies. Adversarial training ~\cite{dong2023enemy,liu2025parameter} strengthens model robustness by training with adversarial examples, enabling them to learn more resilient decision boundaries between benign and attack images. 

\begin{figure}[t] % 这四个字母可以出现一个或多个：htbp 代表图片插入位置的设置
    \centering % 图片居中
    {
    \includegraphics[width=1\linewidth]{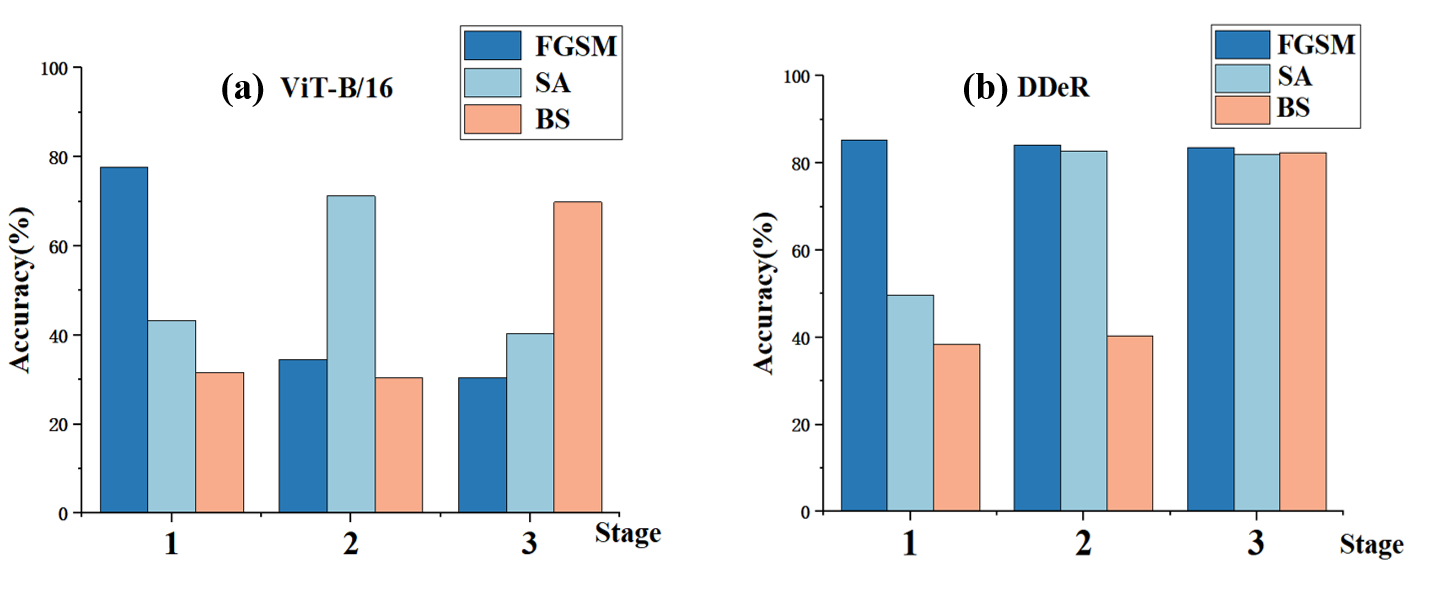}}
    \vspace{-0.2in}
    \caption{Perform continuous adversarial training (CAT) against attack sequence [FGSM~\cite{goodfellow2014explaining}, Square Attack (SA)~\cite{andriushchenko2020square}, BruSLeAttack (BS)~\cite{vo2024brusleattack}]. At each training stage, we evaluate the robust accuracy of (a) baseline ViT-B/16 and (b) our method on the CIFAR-10 dataset against all attacks.}
    \label{fig:teaser}
     \vspace{-0.2in}
\end{figure}

Despite the effectiveness of adversarial training, the ongoing evolution of adversarial attacks presents a continuous game of strategy between attackers and defenders. As new attack methods emerge, previously adversarially trained models may become vulnerable again, thus raising a critical question of `\textbf{\textit{whether the adversarial training can be continuously applied to counteract an ever-expanding range of adversarial attacks?}}' To answer this, we conduct a pilot study in Fig.~\ref{fig:teaser}(a). We find that while adversarial training initially improves the robustness of the model against current attacks, its defense against past attacks gradually weakens as training continues. This suggests that continuously applying adversarial training faces the well-known issue of catastrophic forgetting, which reveals the limitation of static or one-time adversarial training.

In this paper, we thus focus on the task of \textbf{Continual Adversarial Training (CAT)}, aiming to help models adapt to new adversarial attacks over time while maintaining their robustness. The challenge is to ensure that the model learns to defend against new attacks without forgetting previous defenses or causing knowledge conflicts.
A straightforward approach is to apply methods from traditional continual learning~\cite{wang2024comprehensive}. For example, some works~\cite{wang2024sustainable,wang2023continual} store previously used training data and replay it during training to retain defense knowledge. However, this raises privacy concerns. Other studies~\cite{zhou2024defense} use data augmentation and knowledge distillation to refine decision boundaries, but often lead to performance trade-offs.
Moreover, all these approaches rely on a common set of model parameters for all attack sequences. As a result, the model struggles to balance different defense strategies effectively, leading to suboptimal performance.

To this end, we propose a novel framework, called Dual-level Defense Routing (DDeR), for continual adversarial training.
Draw inspiration from MoE~\cite{mustafa2022multimodal}, our first-level defense routing contains a set of defense experts and routers. The router can automatically select several appropriate experts to process the input, minimizing the negative effects of adversarial noise hidden in the image.
This level highlights the significance of knowledge sharing. Although the set of experts is the same for all kinds of adversarial attacks, including past and potential future attacks, the robustness of models against diverse adversarial attacks can be effectively strengthened by integrating the knowledge of different experts. 
Our second-level defense routing is guided by an Adversarial Sentinel Network (ASN), a sub-network that operates independently of the main model. The role of ASN is to assign a suitable router to the incoming image. The number of routers is designed to grow gradually as continual adversarial training progresses. Such an incremental design of independent structures effectively mitigates the limitations of knowledge forgetting or conflict that arise in the parameter-sharing model, as shown in Fig.~\ref{fig:teaser}(b).
Meanwhile, each router is instantiated by a fully connected layer, the increase in the number of parameters is thus minimal.

The purpose of ASN is not to recognize the specific category of adversarial perturbations, but to determine which router should process the input. 
However, since routers are trained independently and unaware of each other, and attack types are unknown at inference, the ASN remains disabled during inference.
To address this, we further introduce a Pseudo-task Substitution Training (PST) strategy, which leverages the distribution discrepancy in image data to establish correlations between routers. More concretely, we first store the mean and covariance of each dataset used in continual adversarial training. Then, we use the re-parameterization technique \cite{wang2019implicit} to sample training images and train ASN to classify which stage of continual adversarial training the input image belongs to. It can not only tackle the problem of ASN training/testing but also eliminate the need to store large amounts of historical training data, thereby avoiding privacy issues. Our contributions can be summarized as follows,

\begin{itemize}
\item We propose a Dynamic Dual-Level Defense Routing framework for Continual Adversarial Training, which autonomously selects appropriate routers to integrate knowledge from different specialized experts, ensuring effective feature representation and enhancing the model's resilience against evolving adversarial attacks.
\item We propose a Pseudo-task Substitution Training strategy to assist the Adversarial Sentinel Network in selecting an appropriate router for processing inputs. This strategy leverages the data distribution discrepancy to establish correlations between routers, reducing historical data storage requirements and mitigating privacy risks.
\item Our method achieves the SOTA defense performance over multiple attack sequences among CIFAR-10, CIFAR-100, and ImageNet-1K datasets, maintaining high accuracy on clean samples and unseen corruption attacks. This demonstrates its sustained robustness against evolving attack methods.
\end{itemize}

\begin{figure*}[t] % 这四个字母可以出现一个或多个：htbp 代表图片插入位置的设置
    \centering % 图片居中
    % \resizebox{1\linewidth}{!}{
    \includegraphics[width=0.95\linewidth]{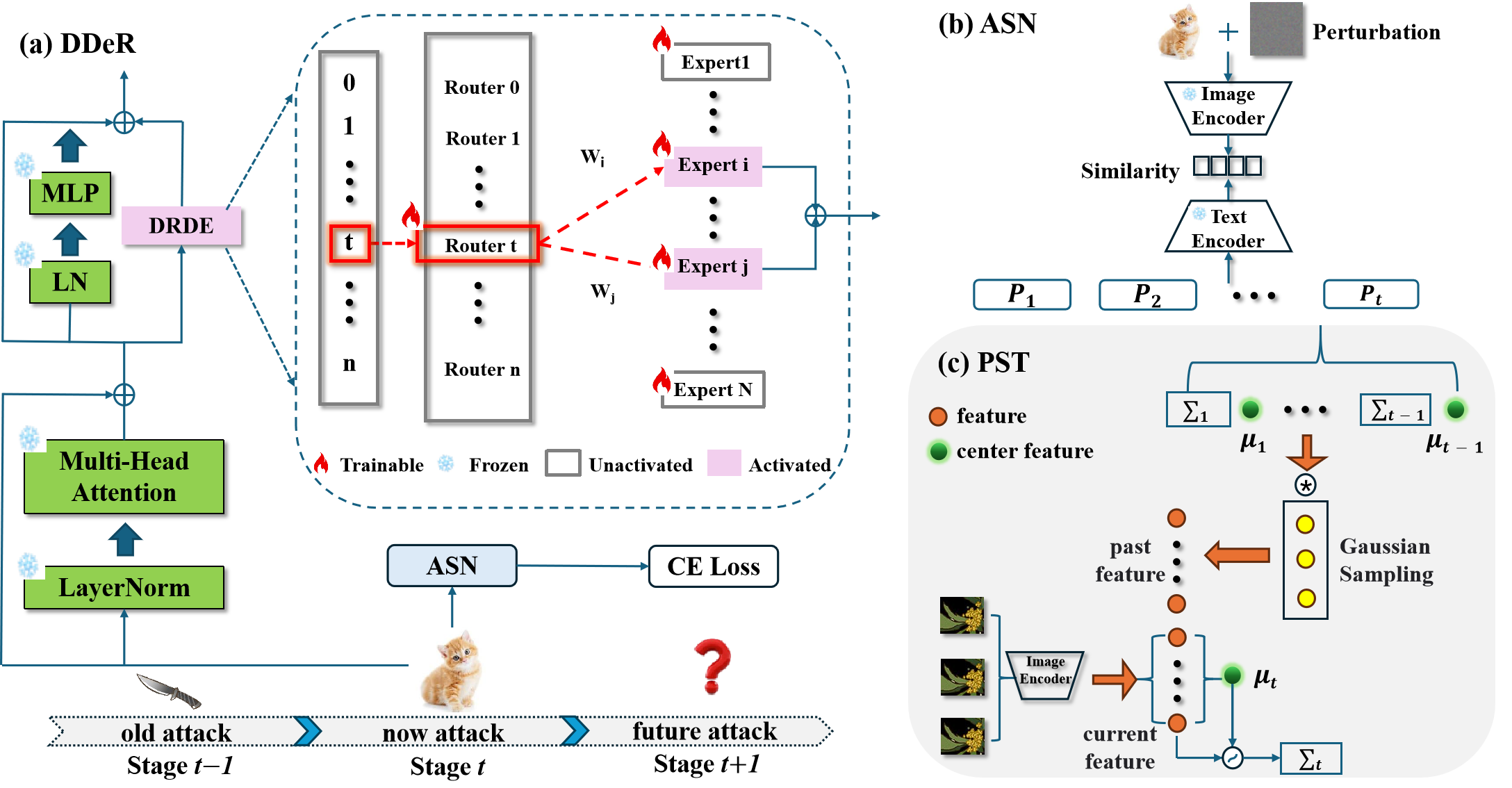}
     \vspace{-0.1in}
    \caption{(a) Schematic diagram of DDeR. During training, each attack type is assigned to a dedicated router, which processes image features from the multi-head attention mechanism and generates expert weights to aggregate multiple expert outputs. (b) The training process of ASN. We represent textual information for each attack type using context vectors and class word embeddings, which are fed into a text encoder to compute similarity scores with input image features. The context vectors are then updated via cross-entropy loss. (c) Autonomous Router Selection via PST. We propose a feature-space-enhanced PST strategy that utilizes distributional variations in image data to establish correlations between routers, ensuring the accurate assignment of images to the appropriate router.
    }
    \label{fig:framework}
    \vspace{-0.15in}
\end{figure*}

\section{Related Work}
% \label{sec:related work}
\noindent\textbf{Adversarial Attacks.}
Adversarial attacks~\cite{li2025enhancing,li2023sibling} mislead deep learning models by adding subtle perturbations to the input. Based on the attacker's access to model information, these attacks are categorized into white-box and black-box types. White-box attacks~\cite{carlini2017towards,xu2025stealthy} assume full knowledge of the model’s architecture, parameters, and gradients, and typically generate perturbations using gradient-based, optimization-based, or decision-boundary-based methods. Though highly effective, their practicality is limited. In contrast, black-box attacks~\cite{yang2024mma,jia2022adv,park2025mind,ren2025improving} require no internal model knowledge and rely solely on input-output behavior, often leveraging adversarial transferability or query-based strategies. As attack techniques continue to evolve, developing robust and sustainable defense mechanisms has become an urgent priority.

\noindent\textbf{Adversarial Training.}
Adversarial training~\cite{madry2017towards}, which enhances model robustness by incorporating adversarial examples during training, has become a leading defense strategy. Recent approaches improve performance by introducing regularization~\cite{li2025pbcat}, knowledge distillation~\cite{levi2025kdat}, and loss-smoothing techniques~\cite{wei2025identifying}.AUTE~\cite{huang2025aute} combines ensemble models with peer-alignment and self-unlearning techniques to effectively enhance adversarial robustness during training. However, adversarial training often sacrifices standard accuracy, prompting research on balancing robustness and clean accuracy. For instance, soft labels and logit-based regularization frameworks~\cite{yin2024boosting} are developed to address this trade-off. Nonetheless, large domain gaps between adversarial samples and the issue of catastrophic forgetting in defense models continue to hinder adaptation to evolving attacks and the realization of sustained robustness.

\noindent\textbf{Continuous Adversarial Defense.}
As attacks are iteratively updated, researchers begin to focus on how to achieve a defense mechanism that continuously expands the attack sequence through adversarial training. Recent studies propose various solutions from the perspective of memory replay. \cite{zhou2024defense} proposes an Anisotropic and Isotropic Replay-based Lifelong Defense (AIR), which constructs pseudo-replay data through data augmentation to mitigate the model's forgetting of past samples. However, given the wide variety of attack types, AIR, a data augmentation-based approach, may fail when there are significant differences between different types of samples. \cite{wang2024sustainable} proposes a Sustainable Self-Evolving Adversarial Training (SSEAT), which involves setting up a buffer to store past samples and utilizing knowledge distillation to effectively balance the model's accuracy on clean samples while defending against different adversarial samples. However, these works have high data storage requirements and privacy issues.

\section{Methodology}

% \label{sec:method}
\subsection{Task Definition and Framework Overview}
\noindent\textbf{Task Definition.}
To develop a defense algorithm adaptable to evolving attacks, we address the Continuous Adversarial Training (CAT) task. Analogous to task-incremental learning, the model sequentially learns a series of attacks $\mathcal{A}=\{A^0, A^1,\ldots,A^t,\ldots\}$, where $A^t$ denotes the $t_{th}$ type of attack and $A^0$ represents clean samples. At $t_{th}$ stage, the training dataset  ${A}^t=\{x_i^t,y_i^t\}_{i=1}^{N^t}$ comprises adversarial inputs $x_i^t$, their labels $y_i^t$, and dataset size $N^t$. The attack type $t$ is provided during training to guide robustness learning, but is not available during inference, aligning with real-world deployment scenarios. Evaluation is performed on samples from all attack types.

\noindent\textbf{Framework Overview.}
As shown in Fig.~\ref{fig:framework}(a), we propose a Dynamic Dual-level Defense Routing (DDeR), a novel framework designed to enhance defense adaptability against evolving attack sequences through a dynamically scalable architecture, addressing the limitations of shared model parameters.  
Our first-layer defense routing is inspired by MoE~\cite{mustafa2022multimodal} and built upon a frozen CLIP model~\cite{radford2021learning}. It integrates a set of defense experts and a router to form Dynamic Routing of Defensive Experts (DRDE). The router autonomously selects a combination of optimal experts, implemented as lightweight LoRA~\cite{hu2021lora} modules, to process inputs, promoting efficient knowledge sharing and specialization.  
The second-level defense routing is managed by the Adversarial Sentinel Network (ASN), which operates independently to assign appropriate routers to incoming images. To ensure ASN remains effective during inference despite routers learning independently during training, we introduce Pseudo-task Substitution Training (PST), which exploits distributional discrepancies in image data to establish inter-router correlations. 

\subsection{Dynamic Routing of Defensive Experts}
To cope with the continuous emergence of novel attack types and the problem of suboptimal defense performance caused by all samples sharing the same set of model parameters, we propose a Dynamic Routing of Defense Experts (DRDE) module, which is an extensible architecture based on MoE~\cite{mustafa2022multimodal}. DRDE continuously expands the router as new attack types increase, aggregating the outputs of defense experts for each attack type to form a specific result, thereby achieving adaptive and optimized defense strategies. For attack type $t$, DRDE contains a set of defense experts $\{M_i\}_{i=1}^{n}$ and a data-sensitive router $\mathcal{R}^t$, where $n$ denotes the number of experts.

To ensure scalability and efficiency, we employ LoRA~\cite{hu2021lora} as lightweight defense experts within DRDE, which decomposes frozen model parameters into a low-rank trainable space, allowing efficient adaptation to diverse attacks with minimal computational overhead. Each router is implemented using a single fully connected layer, preventing an excessive increase in model parameters as the number of routers grows during continual adversarial training. This effectively mitigates issues such as knowledge forgetting and conflicts in parameter-sharing models.

\noindent\textbf{Defensive Experts.}
In DRDE, the data-sensitive router $\mathcal{R}^t$ dynamically selects and activates multiple defense experts $\{M_i\}_{i=1}^{n}$ to generate adaptive responses for each attack type $t$. The corresponding output $f(x_i^t)$ from experts is computed as:
\begin{equation}               f(x_i^t)=\sum_{i=1}^{n}W_i^t{M}_i(\mathbf{x}_i^t)
\end{equation}
where $W^{t}=\{W_{i}^{t}\}_{i=1}^{n}$ indicates the gating weights assigned by $\mathcal{R}^t$, dictating each expert's contribution, and $x_i^t$ represents the input image for attack type $t$. 

Given the distinct data distributions of adversarial samples generated by different attack methods, the contribution of each expert should adapt dynamically to varying attack types. To achieve this, we incorporate image features $\mathbf{f}$ into the router, enabling adaptive weight assignment to defense experts. The gating weights are computed as follows:
\begin{equation}
W^t=Softmax(topk(\mathcal{R}^t(\mathbf{f})))
\end{equation}
where $\mathcal{R}^t$ projects image features $\mathbf{f}$ onto a 1-D vector, quantifying each expert's contribution. The $topk(\cdot)$ selects the top $k$ most influential experts by assigning their corresponding weights to them while setting $-\infty$ for the remaining experts. Finally, $Softmax(\cdot)$ is applied to normalize the selected weights, achieving optimal expert aggregation.

\noindent\textbf{Dynamic Expert Updating.}
To adapt the model to emerging attacks, we introduce a Dynamic Expert Updating (DEU) strategy, which facilitates the addition of new routing modules and updates to defense expert parameters when a novel attack is encountered. Defense experts are trained via standard backpropagation, guided by a momentum-based fusion strategy to retain past knowledge and foster cross-type collaboration, thus mitigating catastrophic forgetting and enhancing robustness.  
Specifically, after training on previous data, the top $k$ most active expert parameters are stored temporarily to maintain type-specific knowledge. When a new attack emerges, these stored parameters are linearly fused with the existing ones, replacing the original expert parameters. The linear fusion is defined as: 
\begin{equation}
    \psi^*=\rho\psi_v+(1-\rho)\psi_u 
\end{equation}
where $\psi^*$ denotes the fused expert parameters, $\psi_v$ and $\psi_u$ represent the temporarily stored expert parameters and the expert parameters from the trained model, respectively. The fusion weight $\rho$ controls the balance between the old and new parameters, which is set to 0.5 to prioritize the retention of knowledge from previous attacks. This allows the router to access stored parameters when encountering new tasks, leveraging past knowledge while optimizing experts for task-specific features, thereby preserving and effectively utilizing historical information.

\subsection{Adversarial Sentinel Network}
In real-world scenarios, the attack type of a sample is unavailable during testing, so the sample cannot be assigned to the appropriate router. To enable autonomous assignment of the most suitable router for each input, we introduce an Adversarial Sentinel Network (ASN), which fully leverages CLIP's prompt learning capability to construct learnable textual vectors for each attack type. These vectors are fed into the text encoder to generate a set of text features. We then compute the similarity between these text features and the image features, which allows the system to automatically match the input to the most relevant router, enabling the router to integrate outputs from the corresponding experts and produce the final result.

Specifically, the textual vectors are composed of a class word vector and context vectors. The class word vector for the attack type $t$ is defined as $\text{TYPE}_t$. Different from using fixed templates to generate context vectors directly from original text, we adopt a set of $M$ learnable vectors $[V]m$ (where $m \in 1, \ldots, M$) to represent the context vectors of attack type $t$. These vectors are defined in the word embedding space, making them learnable and providing greater adaptability. Each vector has the same dimensionality as a word embedding, \emph{i.e.}, 512. The final textual vectors $P_t$ to the text encoder are the concatenation of the context vectors and the class word vector, as shown below:
\begin{equation}
P_t = [V]_1{...}[V]_M[\text{TYPE}_t]
\end{equation}

ASN learns distinct context vectors for each attack type, enabling customized context that facilitates accurate router assignment. We optimize the context parameters using cross-entropy loss. Subsequently, the similarity logits between image and text features are computed to guide router selection and integrate outputs accordingly.

\begin{table*}[t]
\centering{
\setlength{\tabcolsep}{5mm}{
\begin{tabular}{lcccccccc}
\toprule
\multicolumn{1}{c}{\textbf{\textsc{Methods}}} & \textbf{FGSM} & \textbf{BIM} & \textbf{PGD} & \textbf{SA} & \textbf{BS} & \textbf{MCG} & \textbf{DIM} & \textbf{Clean}\tabularnewline
\midrule
PGD-AT & 48.92 & 47.08 & 50.58 & 45.32 & 42.83 & 40.65 & 47.12 & 72.13 \tabularnewline
AWP & 44.50 & 41.89 & 48.12 & 40.93 & 41.36 & 43.82 & 40.21 & 68.77 \tabularnewline
RIFT & 51.65 & 52.91 & 53.89 & 49.21 & 51.72 & 50.45 & 47.91 & 73.54 \tabularnewline
LBGAT & 53.14 & 53.22 & 54.37 & 51.26 & 52.89 & 52.33 & 50.10 & 71.95 \tabularnewline
AFD & 51.12 & 51.45 & 53.14 & 50.83 & 50.65 & 52.78 & 49.56 & 72.41 \tabularnewline
SSEAT & 54.82 & 55.47 & 56.47 & 56.11 & 57.65 & 56.37 & 55.93 & 71.52 \tabularnewline
\midrule
DDeR (\textit{Ours}) & \textbf{60.71} & \textbf{61.23} & \textbf{62.83} & \textbf{61.55} & \textbf{63.07} & \textbf{62.47} & \textbf{64.14} & \textbf{75.92} \tabularnewline
\bottomrule
\end{tabular}}}
\caption{Comparison results under a long attack sequence [FGSM, BIM, PGD, SA, BS, MCG, DIM] on ImageNet-1K.
% \vspace{-0.1in}
\label{Tab:imagenet_1}}
\vspace{-0.15in}
\end{table*}

\subsection{Pseudo-task Substitution Training}
During ASN training, each stage involves only one attack type, resulting in independent routers. However, at inference time, the attack type of a test sample is unknown, and this lack of inter-router coordination hampers accurate router selection, causing ASN failure. To address this, a router capable of adapting to unseen data is required. While data replay can revisit past knowledge to guide router selection in CAT, it incurs substantial storage overhead.

Recent studies \cite{wang2021regularizing, wang2019implicit} show that feature expansion in deep feature space can generate diverse intra-class representations, effectively augmenting data at the feature level. Inspired by this, we propose a Pseudo-task Substitution Training (PST) strategy to mitigate the storage cost of saving original samples. PST leverages the mean and covariance matrix of stored features from previous attack types to expand feature representations, allowing the ASN to incorporate prior data features during router training. This approach establishes correlations between routers while reducing storage requirements.

Specifically, as in Fig.~\ref{fig:framework}(c), in the feature space, we compute the ``center" of each attack type, \emph{i.e.}, the average feature representation. Then, we calculate the variance between the center and feature representations of samples with the same attack type within each mini-batch. These variances are progressively aggregated into a covariance vector using a moving average, which can be formulated as:
\begin{equation}
    \Sigma_t=\mathbb{E}[\left(\boldsymbol{\mu}_{t}-\boldsymbol{f}_{t}\right)^2]
\end{equation}
Resampling the feature representations of past attack types based on the covariance matrix and mean can be defined as:
\begin{equation}
\boldsymbol{f}_{k}=\mu_{k}+L_kz
\end{equation}
\begin{equation}
    \Sigma_k=L_kL_k^T
\end{equation}
where $\boldsymbol{f}_k$ is the sampled feature from the past attack type $k$. $z$ is a random vector sampled from a standard normal distribution $\mathcal{N}(0,I)$. $L_k$ is the Cholesky decomposition matrix of the covariance matrix $\Sigma_k$. In training new attacks, we sample features from the stored mean and covariance matrices of past attack types to generate a subset of features. This process enriches the diversity of features and helps form correlations between routers to achieve input-based router adaptive selection during the inference phase.

\section{Experimental}
\subsection{Experimental Setting}
\noindent \textbf{Datasets.}
We evaluate our method on ImageNet-1k~\cite{deng2009ImageNet}, which is a widely used benchmark in adversarial attack and defense research. The model sequentially encounters a series of attacks, with training and testing data generated according to the corresponding attack methods. The model is trained on the generated adversarial examples over the training set, while the test set is exclusively used to assess its robustness. \textit{More experimental results over CIFAR-10 and CIFAR-100~\cite{krizhevsky2009learning} can be found in the Supplementary Material.}

\noindent \textbf{Attack Algorithms.}
We assess the robustness of all baselines on ImageNet-1k against various strong adversarial attacks. 
(1) For Fast Gradient Sign Method (FGSM)~\cite{goodfellow2014explaining} with perturbation size $\epsilon=8/255$. 
For Projected Gradient Descent (PGD)~\cite{madry2017towards} with $\epsilon=8/255$, step size as 2/255, and the steps as 20. 
For CW attack ~\cite{carlini2017towards} with steps as 1000 and learning rate as 0.01. For DIM~\cite{wu2021improving} with $\epsilon=16/255$ 16/255, steps as 10, and weight parameters as 1.0. 
For DeepFool (Df)~\cite{moosavi2016deepfool} with steps as 50 and an overshoot of 0.02. 
For AutoAttack (AA)~\cite{croce2020reliable} with $\epsilon$ as 8/255.
For Square Attack (SA)~\cite{andriushchenko2020square} with the number of queries as 10,000 and $\epsilon=0.05$. 
For BruSLeAttack (BS)~\cite{vo2024brusleattack} with the queries as 1000 and the sparsity as 0.2\%. 
For MCG~\cite{yin2022mcg} with the number of queries as 10,000 and $\epsilon=0.0325$.
(2) For the unseen adversaries, we follow the implementation of \cite{jiang2024ramp}, setting $\epsilon=12$ for the fog attack, $\epsilon=0.5$ for the snow attack, $\epsilon=60$ for the Gabor attack, $\epsilon=0.125$ for the elastic attack, and $\epsilon=0.125$ for the jpeglinf attack with 100 iterations.
(3) For attacks with different norms, similar to~\cite{croce2022adversarial}, we use APGD with 5 steps for $l_\infty$ and $l_2$ attacks and $15$ steps for $l_1$ attacks. The perturbation budgets are set to $\epsilon_1=12,\epsilon_2=0.5,\epsilon_\infty=8/255$.

\noindent \textbf{Implementation Details.}
We conduct experiments on a single NVIDIA RTX 3090 GPU with PyTorch. For ImageNet-1K, images are resized to 224×224 and augmented using horizontal flipping and random cropping. The model architecture is ViT-B/16. For training, we use a batch size of 64 and optimize the model using the Adam optimizer with a learning rate of $1 \times 10^{-3}$. The router and expert parameters are trained for 30 epochs. During ASN training, a context vector of length 16 was learned over 20 epochs.

\noindent \textbf{Competitors.}
We compare with traditional adversarial training, such as PGD-AT \cite{madry2017towards}, AWP \cite{wu2020adversarial}, RIFT \cite{zhu2023improving}, LBGAT~\cite{cui2021learnable} and AFD~\cite{zhou2025mitigating}.
Meanwhile, we compare with CAT method SSEAT~\cite{wang2024sustainable} under the same settings. For fairness, traditional adversarial training methods are further fine-tuned on adversarial attacks, with knowledge distillation~\cite{hinton2014distilling} applied between old and updated models to mitigate catastrophic forgetting. All competitors use the ViT-B/16 model.

\noindent \textbf{Evaluation Metrics.}
We set different attack sequences, including [FGSM, PGD, CW, AA, Df], [FGSM, BIM, PGD, SA, BS, MCG, DIM], and [$L_\infty$, $L_2$, $L_1$]. After CAT models undergo continual training following the attack sequence and the traditional adversarial training methods complete their respective training processes, we evaluate their robustness against all attacks in the sequence, as well as their classification accuracy on clean test data.

\begin{table}[t]
% \small
\begin{centering}
% \vspace{-0.15in}
\setlength{\tabcolsep}{1mm}{
\begin{tabular}{lcccccc}
\toprule
\multicolumn{1}{c}{\textbf{\textsc{Methods}}} & \textbf{FGSM} & \textbf{PGD} & \textbf{CW} & \textbf{AA} & \textbf{Df} & \textbf{Clean} \tabularnewline
\midrule
% \midrule
PGD-AT   & 46.71 & 37.44 & 38.08 & 37.66 & 38.97 & 60.12 \tabularnewline
AWP      & 49.85 & 40.21 & 41.12 & 40.46 & 40.38 & 64.91 \tabularnewline
RIFT     & 51.82 & 47.37 & 46.31 & 46.22 & 48.09 & 62.55 \tabularnewline
LBGAT    & 48.76 & 44.59 & 45.07 & 44.38 & 45.66 & 61.73 \tabularnewline
AFD      & 52.03 & 53.11 & 53.89 & 53.27 & 54.42 & 66.18 \tabularnewline
SSEAT    & 56.84 & 57.36 & 58.75 & 59.41 & 59.07 & 61.01 \tabularnewline
\midrule
DDeR (\textit{Ours}) & \textbf{68.56} & \textbf{67.38} & \textbf{68.53} & \textbf{68.49} & \textbf{69.79} & \textbf{70.34} \tabularnewline
\bottomrule
\end{tabular}}
\par\end{centering}
\caption{Comparison results under a short attack sequence [FGSM, PGD, CW, AA, Df] on the ImageNet-1K.
% \vspace{-0.1in}
\label{Tab:imagenet_2}}
% \vspace{-0.15in}
\end{table}

\begin{table}[t]
\begin{centering}
\setlength{\tabcolsep}{1.3mm}
% \scriptsize
% \small
\begin{tabular}{lccccc}
\toprule
\multicolumn{1}{c}{\textbf{\textsc{Methods}}} & \textbf{$L_\infty$} & \textbf{$L_2$} & \textbf{$L_1$} & \textbf{UNION} & \textbf{CLEAN} \tabularnewline
\midrule
PGD-AT     & 51.83 & 36.11 & 29.29 & 29.29 & 74.12 \tabularnewline
AWP        & 48.24 & 36.40 & 28.05 & 28.05 & 71.12 \tabularnewline
LBGAT      & 52.97 & 38.02 & 31.41 & 31.41 & 72.83 \tabularnewline
RIFT       & 56.41 & 40.52 & 34.21 & 34.21 & 74.03 \tabularnewline
AFD        & 54.09 & 39.70 & 33.75 & 33.75 & 73.51 \tabularnewline
SSEAT      & 56.80 & 41.19 & 36.94 & 36.94 & 73.89 \tabularnewline
\midrule
DDeR \textit{(Ours)}   & \textbf{58.33} & \textbf{42.75} & \textbf{38.21} & \textbf{38.21} & \textbf{76.84}  \tabularnewline
\bottomrule
\end{tabular}
\par\end{centering}
\caption{Comparison results against adversarial attacks with different norms under the CAT task on ImageNet-1K. 
% \vspace{0.1in}
\label{Tab:Attacks of different norms}}
\vspace{-0.1in}
\end{table}

\begin{table}[t]
% \small
% \vspace{-0.1in}
\begin{centering}
\setlength{\tabcolsep}{1.5mm}{
% \scriptsize
\begin{tabular}{lcccccc}
\toprule
\multicolumn{1}{c}{\textbf{\textsc{Methods}}} & \textbf{fog} & \textbf{snow} & \textbf{gabor} & \textbf{elastic}  & \textbf{jpeglinf}\tabularnewline
\midrule % \midrule
PGD-AT  & 40.24 & 21.37 & 41.87 & 48.14 & 48.56 \tabularnewline
AWP     & 35.78 & 22.34 & 39.20 & 42.89 & 45.37 \tabularnewline
RIFT    & 44.59 & 32.59 & 47.37 & 57.37 & 53.46 \tabularnewline
LBGAT   & 42.15 & 30.28 & 44.73 & 54.18 & 50.22 \tabularnewline
AFD     & 45.36 & 39.25 & 48.67 & 50.66 & 55.48 \tabularnewline
SSEAT   & 44.13 & 35.79 & 49.17 & 51.32 & 58.72 \tabularnewline
\midrule
DDeR (\textit{Ours})  & \textbf{47.49} & \textbf{48.31} & \textbf{53.49} & \textbf{60.45} & \textbf{59.36} \tabularnewline
\bottomrule
\end{tabular}}
\par\end{centering}
\caption{Comparison results about CAT models against unseen natural corruption attacks over ImageNet-1K. 
% \vspace{-0.1in}
\label{Tab:unknown_adversarial_attacks}}
% \vspace{-0.2in}
\end{table}

\begin{table}[t]
% \small
% \vspace{-0.1in}
\begin{centering}
\setlength{\tabcolsep}{1.8mm}{
% \scriptsize
\begin{tabular}{lccc}
\toprule
\multicolumn{1}{c}{\textbf{\textsc{Methods}}} & \textbf{PM (MB)} & \textbf{Param} & \textbf{Cost (GFLOPS)} \tabularnewline
\midrule % \midrule
PGD-AT  & 22,543 &  82M & 38.63  \tabularnewline
AWP & 28,449 & 85M & 45.17 \tabularnewline
 LBGAT& 25,475 & 92M & 48.39 \tabularnewline
 RIFT  & 28,173 & 103M & 41.70 \tabularnewline
 AFD    & 27,801 & 89M & 42.58 \tabularnewline
SSEAT & 32,475 & 83M & 40.37 \tabularnewline
\midrule
DDeR (\textit{Ours})  & \textbf{19,437} & \textbf{58K} & \textbf{23.78} \tabularnewline
\bottomrule
\end{tabular}}
\par\end{centering}
\caption{Comparison over Train-time peak memory allocation (PM), the number of learnable parameters (Param), and the total computational cost (Cost) on ImageNet-1K.
% \vspace{-0.1in}
\label{Tab:Attacks cost}}
\vspace{-0.1in}
\end{table}

\subsection{Experimental Results}

We evaluate the defense performance of DDeR against various competitors on several datasets, assessing its robustness across diverse attack methods, adversarial perturbations under different norms, and clean samples. Additionally, we compare the number of learnable parameters and peak memory allocation of ours and competitors.

\noindent \textbf{DDeR effectively realizes the best defense performance against several attack sequences and achieves the highest accuracy over clean samples.} 
As in Tab.~\ref{Tab:imagenet_1} and Tab.~\ref{Tab:imagenet_2}, under various attack sequences, our DDeR bests all competitors with large margins over both clean and adversarial samples. 
Besides, we evaluate robustness under attack sequences that differ in perturbation norms as shown in Tab.~\ref{Tab:Attacks of different norms}. Consistent with prior findings~\cite{tramer2019adversarial}, our results confirm that improving robustness to one perturbation may inadvertently increase susceptibility to others. Nevertheless, DDeR achieves the highest union accuracy~\cite{laidlaw2020perceptual} among all methods, demonstrating superior general robustness.
The above results demonstrate the superiority of DDeR on CAT task, which stems from its optimized selection and combination of expert knowledge, enabling more precise decision boundaries for different inputs.

\noindent \textbf{DDeR achieves superior defense against unseen attacks.}
Trained on attack sequence [FGSM, PGD, CW, AA, Df] using ImageNet-1K, DDeR is evaluated on unseen natural corruptions~\cite{jiang2024ramp}. In Tab.~\ref{Tab:unknown_adversarial_attacks}, DDeR outperforms all competitors across all unseen attacks. This demonstrates its ability to generalize by leveraging knowledge from known attacks for effective router selection and prediction. Additionally, DDeR achieves a superior robustness-accuracy tradeoff, highlighting its universal robustness.

\noindent\textbf{DDeR exhibits lower memory consumption and requires fewer learnable parameters than competitors for the CAT task.}
As in Tab.~\ref{Tab:Attacks cost}, our DDeR requires significantly less memory than rehearsal-based models, which incur additional storage overhead by retaining past samples. In contrast, DDeR dynamically expands a minimal number of parameters during training, storing only essential feature information and a limited set of historical parameters, thereby optimizing memory efficiency and making it more practical.

\begin{table*}[t]
\centering{
% \small
\setlength{\tabcolsep}{4mm}{
\begin{tabular}{ccccccccccc}
\toprule
\multirow{2}{*}{\textbf{\textsc{No.}}} & \multicolumn{3}{c}{\textbf{Components}} & \multicolumn{6}{c}{\textbf{Method}} \\
\cmidrule(lr){2-4} \cmidrule(lr){5-10}
 & DRDE & PST & ASN & \textbf{FGSM} & \textbf{PGD} & \textbf{CW} & \textbf{AA} & \textbf{Df} & \textbf{Clean} \\
\midrule
$a$ & $\times$ & $\times$  & $\times$  & 48.22 & 49.73 & 46.46 & 50.38 & 58.13 & 49.33 \\
$b$  & $\checkmark$ & $\times$ & $\times$  & 54.17 & 55.93 & 55.26 & 56.39 & 62.32 & 54.46\\
$c$  & $\checkmark$ & $\checkmark$  & $\times$  & 56.67 & 57.36 & 56.43 & 57.22 & 63.79 & 56.17 \\
$d$  & $\checkmark$  & $\times$  & $\checkmark$  & 63.91 & 64.63 & 65.68 & 65.47 & 67.73 & 66.46\\
$e$ & $\checkmark$ & $\checkmark$ & $\checkmark$  & \textbf{68.56} & \textbf{67.38} & \textbf{68.53} & \textbf{68.49} & \textbf{69.79} &  \textbf{70.34} \\
\bottomrule
\end{tabular}}}
\caption{The ablation studies results of our proposed components in DDeR on ImageNet-1K.}
% \vspace{-0.1in}
\label{Tab: Ablation Studies}
\vspace{-0.15in}
\end{table*}

\subsection{Ablation Study}
To verify the role of each module in DDeR, we conduct extensive ablation studies on the following variants: 
(a) ``Baseline": conducted on CAT pipeline using the ViT-B/16 model;
(b) ``+DRDE'': building upon (a),  we introduce shared experts and routers across all attack scenarios, along with a momentum-based parameter fusion strategy that incorporates historical parameters to guide the update of current model parameters;
(c) ``+DRDE+PST'': build upon (b), resampling features via using the mean and covariance matrix of feature representations from past samples;
(d) ``+DRDE+ASN'': build upon (b), we incorporate an Adversarial Sentinel Network (ASN) that exploits data distribution differences to improve router assignment. ASN trains context vectors using current-stage data to measure text-image feature similarity, producing logits that guide router selection;
(e) ``+DRDE+PST+ASN (Ours)'': build upon (d), to enhance the training data of ASN, we employ PST to sample historical examples, thereby improving data diversity.

\noindent \textbf{The efficacy of each component in the proposed method.}
As shown in Tab.~\ref{Tab: Ablation Studies}, by comparing the results of different variants, we notice the following observations,
1) According to the results of ($a$), based on the basic continual learning pipeline, the ``Baseline" suffers from severe knowledge forgetting and fails to retain defense performance against prior attacks;
2) Comparison between ($a$) and ($b$): Results show that introducing experts and routers yields modest performance gains across adversarial samples but does not effectively mitigate catastrophic forgetting;
3) Comparison between ($b$) and ($c$): Results indicate that PST effectively resamples critical past feature information, aiding the retention of previously learned robust representations;
4) Comparison between ($b$) and ($d$): Results demonstrate that using multiple routers with expert assignment per sample type mitigates shared parameter suboptimality, enhancing overall performance;
5) Comparison between ($d$) and ($e$): The results show that our model enhances ASN’s routing for inputs by storing only the mean and covariance of each dataset in CAT tasks, significantly reducing storage. Additionally, preserving representative features of each data type also leads to optimal performance.

\begin{figure}[t] % 这四个字母可以出现一个或多个：htbp 代表图片插入位置的设置
    \centering % 图片居中
    \resizebox{1\linewidth}{!}{
    \includegraphics[width=1\linewidth,height=0.55\linewidth]{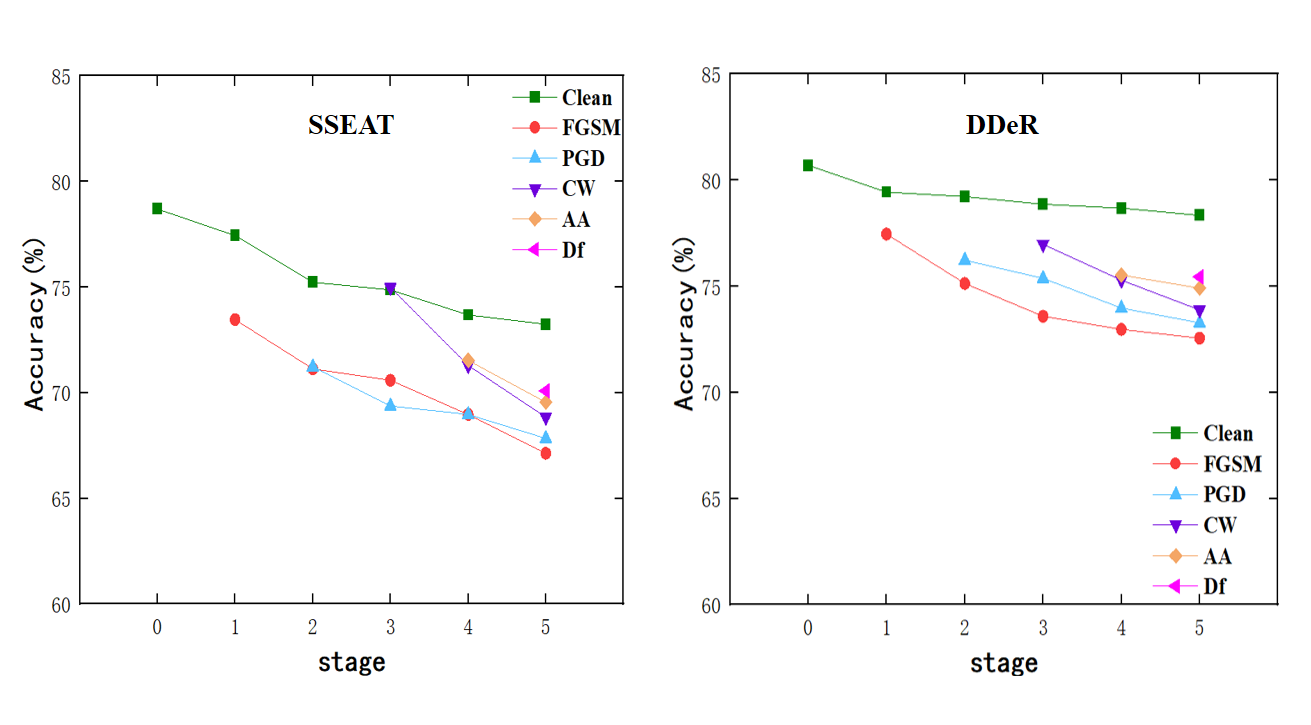}}
    \vspace{-0.2in}
    \caption{The accuracy of SSEAT and DDeR on both current and past attack data at each stage across the attack sequence [FGSM, PGD, CW, AA, Df] on ImageNet-1k.}
    \label{fig:linegraph}
    %\vspace{-0.15cm}
\end{figure}

\begin{figure}[t] % 这四个字母可以出现一个或多个：htbp 代表图片插入位置的设置
    \centering % 图片居中
    \resizebox{1.0\linewidth}{!}{
    \includegraphics[width=1.2\linewidth,height=0.45\linewidth]{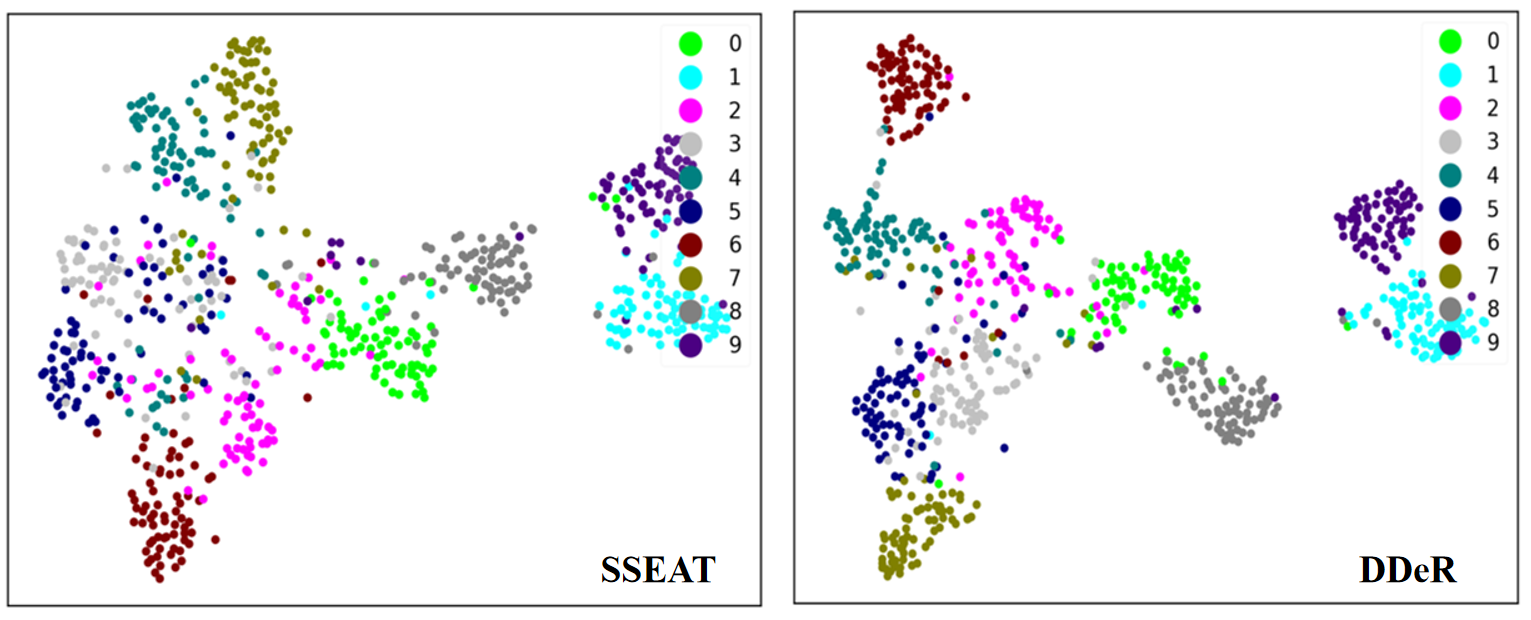}}
    \vspace{-0.1in}
    \caption{Visualizations of feature representations of clean samples by SSEAT and DDeR on ImageNet-1K. Different colors represent clean data of different categories. Zoom in for better views.}
    \label{fig:tsne}
    \vspace{-0.1in}
\end{figure}

\begin{table}[t]
\begin{centering}
\setlength{\tabcolsep}{1.2mm}{
\begin{tabular}{lccccc}
\toprule
\multicolumn{1}{c}{\textbf{\textsc{Methods}}} & \textbf{PGD$_{S}^{50}$} & \textbf{PGD$_{S}^{20}$} & \textbf{PGD$_{W}^{50}$} & \textbf{PGD$_{W}^{20}$} & \textbf{Clean} \tabularnewline
\midrule
PGD-AT & 22.71 & 25.84 & 47.95 & 52.31 & 60.42 \tabularnewline
AWP    & 19.34 & 23.06 & 42.88 & 47.12 & 57.89 \tabularnewline
RIFT   & 21.86 & 24.57 & 46.33 & 50.84 & 60.11 \tabularnewline
LBGAT  & 22.05 & 25.14 & 47.08 & 51.25 & 59.78 \tabularnewline
AFD    & 19.22 & 22.39 & 44.75 & 48.92 & 58.64 \tabularnewline
SSEAT  & 21.76 & 26.10 & 45.88 & 49.53 & 56.71 \tabularnewline
\midrule
DDeR (\textit{Ours}) & \textbf{24.12} & \textbf{27.66} & \textbf{51.03} & \textbf{55.78} & \textbf{62.30} \tabularnewline
\bottomrule
\end{tabular}}
\par\end{centering}
\caption{Defense results with various attack strengths on ImageNet-1K. The $S$ in the lower right corner indicates the strong attack perturbation size as 16/255, and the $W$ indicates the weak attack perturbation size as 8/255.}
\label{Tab:decrease_intensity}
\end{table}

\noindent \textbf{Evaluation under increased attack intensity and iterations.}
Following the model setup in AIR~\cite{zhou2024defense}, the forgetting phenomenon is more severe when progressing “from difficult to easy” tasks. To evaluate model robustness under varying difficulties, we gradually reduce the perturbation magnitude and the number of attack iterations. The CAT model is trained on a sequence of attacks: [$PGD_{S}^{50}$, $PGD_{S}^{20}$, $PGD_{W}^{50}$, $PGD_{W}^{20}$]. As shown in Tab.~\ref{Tab:decrease_intensity}, DDeR consistently outperforms all competitors across all attack settings, achieving robust performance while maintaining high clean accuracy.

\noindent \textbf{Visualization results.}
1) As in Fig.~\ref{fig:linegraph}, it shows the accuracy trajectories of SSEAT and DDeR on various adversarial attacks and clean samples during continual training. The results clearly demonstrate that DDeR achieves superior continual robustness and more effectively alleviates catastrophic forgetting.
2) As in Fig.~\ref{fig:tsne}, after training on [FGSM, PGD, CW, AA, Df], we randomly sampled 100 clean test images from 10 ImageNet-1K classes and extracted features via SSEAT and DDeR. t-SNE visualizations show that DDeR produces more discriminative features, with greater inter-class separation and tighter intra-class clustering, reflecting its strong clean-data accuracy.

\section{Conclusion}
We introduce DDeR for the CAT task, leveraging a corresponding router to adaptively select and integrate suitable experts based on input images, thereby enhancing continual adversarial robustness. To enable efficient router selection, we design the ASN module with PST strategy, leveraging distributional variations in image data to establish correlations between routers, to accurately assign images to the appropriate router. Extensive experiments demonstrate that DDeR achieves state-of-the-art continual robustness and classification accuracy of benign images.

\bibliography{aaai25}
\input{repro}
\end{document}